\documentclass[english]{article}
\usepackage[T1]{fontenc}
\usepackage[latin9]{inputenc}
\usepackage{verbatim}
\usepackage{calc}
\usepackage{array}
\usepackage{mathtools}
\usepackage{multirow}
\usepackage{amsmath}
\usepackage{graphicx}
\usepackage{wasysym}
\usepackage{float}
\usepackage{dsfont}
\usepackage{extarrows}
\usepackage{cite}

\makeatletter

\providecommand{\tabularnewline}{\\}

\usepackage{amsfonts,amssymb}
\usepackage{latexsym}
\usepackage{epsfig}
\usepackage{cite}
\usepackage[colorlinks,linkcolor=blue]{hyperref}
\newcommand{\be}{\begin{equation}}
\newcommand{\ee}{\end{equation}}

\makeatother

\usepackage{babel}
\begin{document}
{}~ \hfill\vbox{\hbox{CTP-SCU/2023004}}\break
\vskip 3.0cm
\centerline{\Large \bf  Timelike entanglement entropy and  $T\bar{T}$ deformation }

\vspace*{10.0ex}
\centerline{\large Xin Jiang$^a$, Peng Wang$^a$, Houwen Wu$^{a,b}$ and Haitang Yang$^a$}
\vspace*{7.0ex}
\vspace*{4.0ex}
\centerline{\large \it $^a$College of Physics}
\centerline{\large \it Sichuan University}
\centerline{\large \it Chengdu, 610065, China} \vspace*{1.0ex}
\vspace*{4.0ex}
\centerline{\large \it $^b$DAMTP, Centre for Mathematical Sciences}
\centerline{\large \it University of Cambridge}
\centerline{\large \it Cambridge, CB3 0WA, UK} \vspace*{1.0ex}
\vspace*{4.0ex}

\centerline{domoki@stu.scu.edu.cn, pengw@scu.edu.cn, hw598@damtp.cam.ac.uk, hyanga@scu.edu.cn}
\vspace*{10.0ex}
\centerline{\bf Abstract} \bigskip \smallskip

In a previous work  \cite{Wang:2018jva} about the  $T\bar{T}$ deformed CFT$_2$,
from the consistency requirement 
of the entanglement entropy, we found that in addition to the usual spacelike entanglement entropy, a timelike entanglement entropy must be introduced and treated equally. 
Motivated by the recent explicit constructions of the timelike entanglement entropy and its bulk dual, we provide a comprehensive analysis of the timelike and  spacelike entanglement entropies in the  $T\bar{T}$ deformed finite size system and finite temperature system. The results confirm our prediction that in the finite size system only the timelike entanglement entropy receives a  correction, while  in the finite temperature system only the usual spacelike entanglement entropy gets a correction. These findings affirm the necessity of  a complete measure including both spacelike and timelike entanglement entropies.

%The purpose of this paper is to provide a comprehensive analysis of the timelike and  spacelike entanglement entropies in the  $T\bar{T}$ deformed finite size system and finite temperature system. The results confirm our previous prediction that in the finite size system only the timelike entanglement entropy receives a  correction, while  in the finite temperature system only the usual spacelike entanglement entropy gets a correction. These findings affirm the necessity of  a complete measure, called general entanglement entropy, for characterizing deformed systems from the quantum information perspective.
%
%
%In this paper, we present a comprehensive analysis of the timelike and traditional spacelike entanglement entropy of a single interval in two distinct deformed conformal field theories. Our results demonstrate that the timelike entanglement entropy exhibits a leading correction due to the deformation in the finite size system, while the standard entanglement entropy only shows a leading correction in the presence of the deformation in the finite temperature system. These findings underscore the importance of a complete measure, called general entanglement entropy, for characterizing deformed systems from a quantum information standpoint.

\vfill 
\eject
\baselineskip=16pt
\vspace*{10.0ex}

\tableofcontents{}

{}

{}

\section{Introduction}

\label{sec:Introduction}

Conformal field theories (CFTs) could be deformed by relevant, marginal
and irrelevant deformations. Relevant and marginal deformations   have been
well studied. Though the irrelevant deformations are non-renormalizable and 
lead no consequence in the IR region, in two-dimensional spacetime, 
they turn out to be generally under control and even solvable for some
particular models. One such solvable irrelevant
deformation is the $T\bar{T}$ deformation \cite{Zamolodchikov:2004ce,Cavaglia:2016oda,Smirnov:2016lqw},
obtained by turning on a $T\bar{T}$ coupling term

\begin{equation}
\frac{dI_{\text{QFT}}^{\left(\mu\right)}}{d\mu}=\int d^{2}x\,\left(T\bar{T}\right)_{\mu},
\end{equation}
where the deformation parameter $\mu$ has the $\left[\text{Length}\right]^{2}$
dimension and $\left(T\bar{T}\right)_{\mu}$is
defined by the stress tensor of the deformed theory. At the leading
order, the deformed theory is   given by
\begin{equation}
I_{\text{QFT}}^{\left(\mu\right)}=I_{\text{CFT}}+\mu\int d^{2}x\,\left(T\bar{T}\right)_{\mu=0} + {\mathcal O}(\mu^2),
\label{eq:action}
\end{equation}
where  $(T\bar{T})_{\mu=0}=\frac{1}{8}\left[T^{\alpha\beta}T_{\alpha\beta}-\left(T_{\alpha}^{\alpha}\right)^{2}\right]$.
It is clear that the Lorentz symmetry is preserved but the conformal
symmetry is broken in the $T\bar{T}$ deformed theory.

The importance of this model is revived partially by the proposition, given 
in Ref.\cite{McGough:2016lol}, that the $\text{AdS}_{3}$ gravity
with a Dirichlet boundary at a finite radial distance $r_{c}$, 
is dual to the $T\bar{T}$ deformed $\text{CFT}_{2}$ living on that Dirichlet
boundary. This nontrivial extension of AdS/CFT correspondence \cite{Maldacena:1997re} is called 
cutoff-AdS/$T\bar{T}$-deformed-CFT (cAdS/dCFT)
correspondence. 
%A lot of progresses on this subject have been achieved in recent years[...].

%The $T\bar{T}$ deformation, flowing towards the ultraviolet (UV)
%region, has a significant impact on the UV physics of the system,
%rather than its infrared (IR) physics. 

In the UV region, out of the many calculable deformed physical
quantities, a particularly important one is the  \emph{entanglement entropy} (EE). 
Until recently, the  entanglement entropy is defined only for spacelike
intervals. Thereinafter, we will refer the spacelike EE to the usual standard EE. Some progresses on the spacelike EE in the $T\bar{T}$ deformed $\text{CFT}_{2}$ have been achieved in recent years \cite{He:2019vzf,He:2020qcs,He:2022xkh}.

With the replica trick \cite{Calabrese:2004eu,Calabrese:2009qy}, in Ref.\cite{Chen:2018eqk},  
the $T\bar{T}$ deformed  spacelike  EE was calculated  perturbatively for the cylindrical topology.
Intriguingly, the  $T\bar{T}$ correction to the spacelike  EE 
is dependent on  different interpretations of the identical  topology.
When treat the system as a finite size one, there is no leading correction. But the finite temperature
interpretation does receive a leading correction. 
This indicates  that the $T\bar{T}$ correction  
to the spacelike EE can be observed
in the finite temperature system but is invisible in the finite size system.
This result obviously conflicts with the fact that the entanglement entropy is a topological quantity. 
Moreover, without a correction presented
in the finite size system, how do we distinguish between the 
undeformed CFT and the deformed CFT by the  entanglement entropy?

To resolve this inconsistency, in a previous paper \cite{Wang:2018jva}, 
by noticing the fact that the finite size system and the finite temperature system
share the same cylindrical topology under exchanging $t \leftrightarrow x$,
we proposed that in addition to the spacelike  EE, there 
should exist a timelike  EE for  timelike intervals, and  the timelike EE should 
be treated on the same footing as the spacelike EE.
We further predicted, as shown in
Table \ref{tab:correction}, the spacelike EE only receives
a correction in the $T\bar{T}$ deformed finite temperature system,
while the timelike EE only receives a correction
in the $T\bar{T}$ deformed finite size system.

\begin{table}[H]
\begin{centering}
\begin{tabular}{|c|c|c|}
\hline
 & spacelike EE & timelike EE\tabularnewline
\hline
Finite size &  & $\checked$\tabularnewline
\hline
Finite temperature & $\checked$ & \tabularnewline
\hline
\end{tabular}
\par\end{centering}
\caption{\label{tab:correction}The symbol $\checked$ marks   a correction to
the entanglement entropy caused by the $T\bar{T}$ deformation. }
\end{table}

Remarkably, such a timelike EE has  been specifically defined 
via analytical continuation 
and the bulk dual has been explicitly provided in  Ref.\cite{Doi:2022iyj,Narayan:2022afv,Doi:2023zaf,Narayan:2023ebn} recently.
Rather than real-valued as the
spacelike EE, the timelike EE
is  a complex-valued quantity. It is suggested 
that the timelike EE needs to be correctly understood
as a pseudoentropy, which is a non-Hermitian generalization of the usual spacelike EE. Dividing
the total system into two subsystems $A$ and $B$, the pseudoentropy
is defined by the von Neumann entropy, 
\begin{equation}
S_{A}=-\mathrm{Tr}\left[\tau_{A}\log\tau_{A}\right],
\end{equation}
of the reduced transition matrix 
\begin{equation}
\tau_{A}=\mathrm{Tr}_{B}\left[\frac{\left|\psi\right\rangle \left\langle \varphi\right|}{\left\langle \varphi\mid\psi\right\rangle }\right].
\end{equation}
Here, $\left|\psi\right\rangle $ and $\left|\varphi\right\rangle $
are two different quantum states in the total Hilbert space that is
factorized as $\mathcal{H}=\mathcal{H}_{A}\otimes\mathcal{H}_{B}$.
As the usual EE, the pseudoentropy could also be captured by the replica method
\cite{Calabrese:2004eu,Calabrese:2009qy} in path integral formalism.
Denoting the manifold corresponding to $\left\langle \varphi\mid\psi\right\rangle $
as $\mathcal{M}_{1}$ and the manifold corresponding to $\mathrm{Tr}_{A}\left(\tau_{A}\right)^{n}$
as $\mathcal{M}_{n}$, the $n$-th pseudo R\'enyi entropy reads
\begin{equation}
S_{A}^{\left(n\right)}=\frac{1}{1-n}\log\left[\frac{Z_{\mathcal{M}_{n}}}{\left(Z_{\mathcal{M}_{1}}\right)^{n}}\right],
\label{eq:Renyi entropy}
\end{equation}
where $Z_{\mathcal{M}}$ is the partition function over the manifold $\mathcal{M}$.
Taking the limit  $n\rightarrow1$ yields the pseudoentropy
\begin{equation}
S_{A}=\lim_{n\rightarrow1}\frac{1}{1-n}\log\left[\frac{Z_{\mathcal{M}_{n}}}{\left(Z_{\mathcal{M}_{1}}\right)^{n}}\right].\label{eq:entropy}
\end{equation}

\begin{figure}[H]
\begin{centering}
\includegraphics[width=0.25\textwidth]{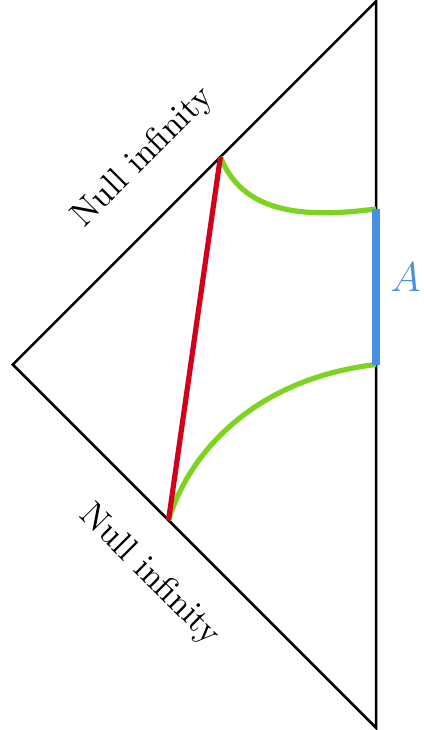}\label{fig:penrose}
\par\end{centering}
\caption{Geodesics connecting to $\partial A$ in the Poincare patch of $\text{AdS}_{3}$.
The red line denotes one timelike geodesic and two green lines denote
two spacelike geodesics. The blue line denotes a timelike interval
$A$ on the conformal boundary.}
\end{figure}

Consider  a two-dimensional CFT in a flat spacetime whose time and space
coordinate are $(t,x)$, and we  now  construct a \textit{general
entanglement entropy}, which contains both spacelike and timelike
components. By choosing a general interval $A=[(t_1,x_1),(t_2,x_2)]$ and using the replica trick, we could obtain the general EE
\begin{equation}
S_A=\frac{c_{\text{AdS}}}{6}\log\left[\frac{\left(x_{1}-x_{2}\right)^{2}-\left(t_{1}-t_{2}\right)^{2}}{\epsilon^{2}}\right].
\label{eq:g_entropy}
\end{equation}
 In this paper, the general EE serves primarily as a convenient formula to group timelike and spacelike EEs together, since the spacelike and timelike EEs are just two different limits of the general EE:
 \begin{equation}
 S_A=\begin{cases}
\frac{c_{\text{AdS}}}{3}\log\frac{\left|x_{1}-x_{2}\right|}{\epsilon}, & t_{1}=t_{2},\\
\frac{c_{\text{AdS}}}{3}\log\frac{\left|t_{1}-t_{2}\right|}{\epsilon}+i\frac{c_{\text{AdS}}\pi}{6}, & x_{1}=x_{2},
\end{cases}
\label{eq:twocase}
\end{equation}
with $\epsilon$ the UV cutoff and $c_{\text{AdS}}=3R_{\text{AdS}}/2G_{N}$
the central charge of $\text{CFT}_{2}$.  We emphasize that equations (\ref{eq:entropy}) and (\ref{eq:twocase}) are specific to the Poincare patch of $\text{AdS}_{3}$ and not a universal formula for the general entanglement entropy.
The definition of general EE can also be extended to finite temperature
CFT and finite size CFT, respectively, as shown in section \ref{sec:Entanglement-entropy-in}. Therefore, for a $2d$ CFT that is dual to the Poincare patch of $\text{AdS}_{3}$,
the timelike EE $S_{A}$ of a timelike interval $A$, whose
width is given by $T=\left|t_{1}-t_{2}\right|$, reads
\begin{equation}
S_{A}=\frac{c_{\text{AdS}}}{3}\log\left(\frac{T}{\epsilon}\right)+i\frac{c_{\text{AdS}}\pi}{6}.
\end{equation}
    It is worthwhile to note that, in CFTs,  a naive analytic continuation from the spacelike EE to timelike one is generally incorrect \cite{Doi:2023zaf}.
According to the Ryu-Takayanagi formula \cite{Ryu:2006bv}, Ref.\cite{Doi:2022iyj,Li:2022tsv}
found that the complex-valued extremal surface consists of one timelike
geodesic and two spacelike geodesics, as shown in Figure \ref{fig:penrose}.
Two spacelike geodesics connect $\partial A$ and null infinities,
respectively. The timelike geodesic connects the endpoints of 
two spacelike geodesics  on null infinities. Furthermore, the
length of the timelike geodesic is equal to the imaginary part
of the timelike EE, while the total length of two spacelike
geodesics is equal to the real part of the timelike EE. Related works on the complex-valued extremal surface were also proposed in \cite{Narayan:2015vda,Narayan:2015oka,Narayan:2016xwq}. 

The purpose of this paper is to confirm the predictions made in \cite{Wang:2018jva}.
In the cAdS/dCFT correspondence, the general
EE could be thought of a complete measure, which
always receives a correction from the $T\bar{T}$ deformation. The
spacelike and timelike entanglement entropies are different
limits of the general entanglement entropy. As a consistent check,
we will show that, the leading $T\bar{T}$ correction to the timelike
EE exists in the finite size system, but vanishes
in the finite temperature system. Utilizing the holographic method,
we will show the physical reason why spacelike (timelike) EE 
only receives a correction in finite temperature (size) system.

%
%We refer to refs.\cite{Gorbenko:2018oov,Jeong:2019ylz,Grieninger:2019zts,Chen:2019mis,Donnelly:2019pie,Grado-White:2020wlb,Chakraborty:2020udr,Khoeini-Moghaddam:2020ymm,Cardona:2022cmh,He:2023xnb}
%for more recent works on the holographic studies of the $T\bar{T}$
%deformed CFT.

The remainder of this paper is outlined as follows. In section \ref{sec:Entanglement-entropy-in},
we show the leading correction of the general entanglement entropy
of the $T\bar{T}$ deformed finite temperature $\text{CFT}_{2}$ and
finite size $\text{CFT}_{2}$, respectively. In section \ref{sec:Gravity-dual},
using RT formula, we show that the cutoff-AdS geodesic leads to a
precise estimation of the general entanglement entropy in $T\bar{T}$
deformed CFT. Section \ref{Sec:Conc} is devoted to our conclusions.

\section{General entanglement entropy in $T\bar{T}$ deformed CFT}

\label{sec:Entanglement-entropy-in}

In this section, we  calculate the leading corrections
to the general EEof the $T\bar{T}$ deformed CFT 
living on the cylindrical manifold. After taking different limits, we derive the corrections to the 
spacelike EE and timelike EE respectively, for the finite temperature system and finite size system.
For a $T\bar{T}$ deformed CFT on  $\mathcal{M}$,
the general EE of some subsystem $A\in\mathcal{M}$
is obtained by the definition of pseudoentropy. Substituting equation (\ref{eq:action})
into equation (\ref{eq:entropy}), one could obtain the leading correction
to $S\left(A\right)$:
\begin{equation}
\delta S\left(A\right)=\lim_{n\rightarrow1}\delta S_{n}\left(A\right),\quad\delta S_{n}\left(A\right)=\frac{n\mu}{1-n}\int_{\mathcal{M}}\left[\left\langle T\bar{T}\right\rangle _{\mathcal{M}}-\left\langle T\bar{T}\right\rangle _{\mathcal{M}_{n}}\right].
\end{equation}

\subsection{Finite temperature}

\label{subsec:Finite-temperature}

Consider a $T\bar{T}$ deformed CFT at the inverse temperature $\beta$.
This theory lives on a cylindrical manifold $\mathcal{M}_{1}$, which
has a non-compact spatial direction $x\in\left(-\infty,\infty\right)$
and compact Euclidean time $\tau\in\left(0,\beta\right)$ with the
periodicity $\tau\sim\tau+\beta$. It is well-known that the two-point
correlation function in the finite temperature CFT is
\begin{eqnarray}
\left\langle \mathcal{O}\left(w,\bar{w}\right)\mathcal{O}\left(0,0\right)\right\rangle  & = & \left[\frac{\beta^{2}}{\pi^{2}}\sinh\left(\frac{\pi w}{\beta}\right)\sinh\left(\frac{\pi\bar{w}}{\beta}\right)\right]^{-2\Delta}
\end{eqnarray}
with the complex coordinate $w=x+i\tau$ and the scaling dimension 
$\Delta$. By the replica trick,
the entanglement entropy of a single interval $A$, which has timelike
width $\tau_{0}$ and spacelike width $x_{0}$, is related to the
two-point function of the twist fields
\begin{eqnarray*}
\frac{Z_{\mathcal{M}_{n}}\left(A\right)}{Z_{\mathcal{M}_{1}}^{n}} & = & \Bigl\langle\Phi^{+}\left(0,0\right)\Phi^{-}\left(x_{0}+i\tau_{0},x_{0}-i\tau_{0}\right)\Bigr\rangle\\
 & = & \left[\frac{\beta^{2}}{\pi^{2}\epsilon^{2}}\sinh\left(\frac{\pi\left(x_{0}+i\tau_{0}\right)}{\beta}\right)\sinh\left(\frac{\pi\left(x_{0}-i\tau_{0}\right)}{\beta}\right)\right]^{-2\Delta_{n}}
\end{eqnarray*}
with the dimension$\Delta_{n}=\frac{c_{\text{AdS}}}{24}\left(n-\frac{1}{n}\right),$
the UV cutoff $\epsilon$ and the central charge $c_{\text{AdS}}$.
Therefore, the entanglement entropy obtained by equation (\ref{eq:entropy})
is
\begin{eqnarray}
S\left(A\right) & = & \frac{c_{\text{AdS}}}{6}\log\left[\frac{\beta^{2}}{\pi^{2}\epsilon^{2}}\sinh\left(\frac{\pi\left(x_{0}+i\tau_{0}\right)}{\beta}\right)\sinh\left(\frac{\pi\left(x_{0}-i\tau_{0}\right)}{\beta}\right)\right]\nonumber \\
 & = & \frac{c_{\text{AdS}}}{3}\log\left[\frac{\beta}{2\pi\epsilon}\sqrt{2\cosh\left(\frac{2\pi x_{0}}{\beta}\right)-2\cosh\left(\frac{2\pi i \tau_{0}}{\beta}\right)}\right].\label{eq:temp-entropy}
\end{eqnarray}
Following Ref.\cite{Chen:2018eqk}, one could also calculate $\left\langle T\bar{T}\right\rangle _{\mathcal{M}_{1}}$
and $\left\langle T\bar{T}\right\rangle _{\mathcal{M}_{n}}$ in
the finite temperature CFT\footnote{Here, we easily generalize the purely spacelike subsystem in Ref.\cite{Chen:2018eqk} to the general one.}:
\begin{equation}
\left\langle T\bar{T}\right\rangle _{\mathcal{M}_{1}}=\left(\frac{c_{\text{AdS}}}{12}\right)^{2}\left(\frac{2\pi^{2}}{\beta^{2}}\right)^{2},
\end{equation}
\begin{equation}
\left\langle T\bar{T}\right\rangle _{\mathcal{M}_{n}}=\left(\frac{c_{\text{AdS}}}{12}\right)^{2}\left[\left(\frac{2\pi^{2}}{\beta^{2}}\right)^{2}-\frac{2\pi^{2}}{\beta^{2}}\left(n-1\right)\left(Q+\bar{Q}\right)+O\left(\left(n-1\right)^{2}\right)\right],
\end{equation}
with the meromorphic function
\[
Q\left(w\right)\coloneqq\frac{\sinh^{2}\left(\frac{\pi\left(x_{0}+i\tau_{0}\right)}{\beta}\right)}{\sinh^{2}\left(\frac{\pi\left(x_{0}+i\tau_{0}-w\right)}{\beta}\right)\sinh^{2}\left(\frac{\pi w}{\beta}\right)}.
\]
Intuitively, $Q\left(w\right)$ has two poles ($w=0$ and $w=x_{0}+i\tau_{0}$)
that correspond to the residues
\begin{eqnarray*}
\mathrm{Res}\left(Q,w=0\right) & = & \frac{2\pi}{\beta}\coth\left(\frac{\pi\left(x_{0}+i\tau_{0}\right)}{\beta}\right),\\
\mathrm{Res}\left(Q,w=x_{0}+i\tau_{0}\right) & = & -\frac{2\pi}{\beta}\coth\left(\frac{\pi\left(x_{0}+i\tau_{0}\right)}{\beta}\right).
\end{eqnarray*}
The leading correction to $S\left(A\right)$ caused by $T\bar{T}$
deformation is thus 
\begin{eqnarray}
\delta S\left(A\right) & = & \lim_{n\rightarrow1}\frac{n\mu}{1-n}\int_{\mathcal{M}_{1}}\left(\frac{c_{\text{AdS}}}{12}\right)^{2}\left[\frac{2\pi^{2}}{\beta^{2}}\left(n-1\right)\left(Q+\bar{Q}\right)+O\left(\left(n-1\right)^{2}\right)\right]\label{eq:TTbar-temp}\\
 & = & -\mu\int_{\mathcal{M}_{1}}\left(\frac{c_{\text{AdS}}}{12}\right)^{2}\frac{2\pi^{2}}{\beta^{2}}\left(Q+\bar{Q}\right),
\end{eqnarray}
which, with the help of Cauchy's residue theorem, could be simplified
as
\begin{eqnarray*}
\delta S\left(A\right) & = & -2\pi\mu\int_{0}^{x_{0}}dx\,\left(\frac{c_{\text{AdS}}}{12}\right)^{2}\frac{2\pi^{2}}{\beta^{2}}\times\left[\mathrm{Res}\left(Q,0\right)+\mathrm{Res}\left(\bar{Q},0\right)\right],\\
 & = & -\mu\frac{\pi^{4}c_{\text{AdS}}^{2}}{18\beta^{3}}x_{0}\left[\coth\left(\frac{\pi\left(x_{0}+i\tau_{0}\right)}{\beta}\right)+\coth\left(\frac{\pi\left(x_{0}-i\tau_{0}\right)}{\beta}\right)\right]\\
 & = & -\mu\frac{\pi^{4}c_{\text{AdS}}^{2}}{9\beta^{3}}\frac{x_{0}\sinh\left(\frac{2\pi x_{0}}{\beta}\right)}{\cosh\left(\frac{2\pi x_{0}}{\beta}\right)-\cosh\left(\frac{2\pi i\tau_{0}}{\beta}\right)}.
\end{eqnarray*}
Via an analytical continuation $\tau=it$, the entanglement entropy
in the finite temperature system, to the leading order of $\mu$,
reads
\begin{equation}
S^{(\mu)}\left(A\right)=\frac{c_{\text{AdS}}}{3}\log\left[\frac{\beta}{2\pi\epsilon}\sqrt{2\cosh\left(\frac{2\pi x_{0}}{\beta}\right)-2\cosh\left(\frac{2\pi t_{0}}{\beta}\right)}\right]-\mu\frac{\pi^{4}c_{\text{AdS}}^{2}}{9\beta^{3}}\frac{x_{0}\sinh\left(\frac{2\pi x_{0}}{\beta}\right)}{\cosh\left(\frac{2\pi x_{0}}{\beta}\right)-\cosh\left(\frac{2\pi t_{0}}{\beta}\right)}.\label{eq:temp-gen-ee}
\end{equation}
By choosing $\left(x_{0},t_{0}\right)=\left(0,T\right)$, one obtains the timelike
entanglement entropy in the $T\bar{T}$ deformed finite temperature system
\begin{equation}
S_{T}^{\left(\mu\right)}=\frac{c_{\text{AdS}}}{3}\log\left[\frac{\beta}{\pi\epsilon}\sinh\left(\frac{\pi T}{\beta}\right)\right]+i\frac{c_{\text{AdS}}\pi}{6}.\label{eq:field-temp-timelike}
\end{equation}
Comparing the above result with the timelike EE in the original finite temperature system \cite{Doi:2023zaf},
\begin{equation}
S_{T}=\frac{c_{\text{AdS}}}{3}\log\left[\frac{\beta}{\pi\epsilon}\sinh\left(\frac{\pi T}{\beta}\right)\right]+i\frac{c_{\text{AdS}}\pi}{6},\label{eq:temp-time}
\end{equation}
it is not surprising that, in the  $T\bar{T}$ deformed finite
temperature CFT$_2$, the timelike entanglement entropy does not receive
a correction from the $T\bar{T}$ deformation. Moreover, it is worthwhile
to note that the imaginary part of the timelike entanglement entropy
originates from the complex logarithmic function. For the particular geometry we are considering, if one wishes to match the imaginary component, one should take the principle branch which has the correct magnitude \cite{Li:2022tsv}.
Similarly, choosing $\left(x_{0},t_{0}\right)=\left(X,0\right)$,
the spacelike entanglement entropy with $T\bar{T}$ correction is
determined: 
\begin{equation}
S_{X}^{\left(\mu\right)}\left(A\right)=\frac{c_{\text{AdS}}}{3}\log\left[\frac{\beta}{\pi\epsilon}\sinh\left(\frac{\pi X}{\beta}\right)\right]-\mu\frac{\pi^{4}c_{\text{AdS}}^{2}}{9\beta^{3}}X\coth\left(\frac{\pi X}{\beta}\right),
\end{equation}
which exactly agrees with the result in \cite{Chen:2018eqk}.

\subsection{Finite size}

\label{subsec:Finite-size}

Now we focus on the $T\bar{T}$ deformed field theory living on a cylindrical manifold $\mathcal{M}_{2}$,
which has a non-compact temporal direction $\tau\in\left(-\infty,\infty\right)$
and a compact spatial direction $x\in\left(0,L\right)$ with the periodicity
$x\sim x+L$. It is important to notice that $\mathcal{M}_{1}$ and $\mathcal{M}_{2}$
have the same  topology $R\times S^{1}$. Therefore, by setting
$\beta=L$ and exchanging $x\leftrightarrow\tau$ in equations (\ref{eq:temp-entropy})
and (\ref{eq:TTbar-temp})
and performing the analytical continuation $\tau=it$,
one easily obtains the entanglement entropy in the finite size system
\begin{equation}
S^{(\mu)}\left(A\right)=\frac{c_{\text{AdS}}}{3}\log\left[\frac{L}{2\pi\epsilon}\sqrt{2\cos\frac{2\pi t_{0}}{L}-2\cos\frac{2\pi x_{0}}{L}}\right]+\mu\frac{\pi^{4}c_{\text{AdS}}^{2}}{9L^{3}}\frac{t_{0}\sin\left(\frac{2\pi t_{0}}{L}\right)}{\cos\left(\frac{2\pi t_{0}}{L}\right)-\cos\left(\frac{2\pi x_{0}}{L}\right)}.
\end{equation}
Setting $\left(x_{0},t_{0}\right)=\left(0,T\right)$,
the timelike entanglement entropy indeed receives a correction from
the $T\bar{T}$ deformation
\begin{equation}
S_{T}^{(\mu)}\left(A\right)=\frac{c_{\text{AdS}}}{3}\log\left[\frac{L}{\pi\epsilon}\sin\left(\frac{\pi T}{L}\right)\right]-\mu\frac{\pi^{4}c_{\text{AdS}}^{2}}{9L^{3}}T\cot\left(\frac{\pi T}{L}\right)+i\frac{c_{\text{AdS}}\pi}{6}.\label{eq:field-size-timelike}
\end{equation}
Meanwhile, as expected, the spacelike entanglement entropy does not
receive a correction
\begin{equation}
S_{X}^{(\mu)}\left(A\right)=\frac{c_{\text{AdS}}}{3}\log\left[\frac{L}{\pi\epsilon}\sin\left(\frac{\pi X}{L}\right)\right].
\end{equation}

\begin{table}[H]
\begin{centering}
\begin{tabular}{|c|l|l|}
\hline 
 & EE of a single interval & Leading $T\bar{T}$ correction\tabularnewline
\hline 
\multirow{2}{*}{Finite size} & Timelike: $S_{T}=\frac{c_{\text{AdS}}}{3}\log\left(\frac{L}{\pi\epsilon}\sin\left(\frac{\pi T}{L}\right)\right)+\frac{ic_{\text{AdS}}\pi}{6}$ & $\delta S_{T}=-\mu\frac{\pi^{4}c_{\text{AdS}}^{2}}{9L^{3}}T\cot\left(\frac{\pi T}{L}\right)$\tabularnewline
\cline{2-3} \cline{3-3} 
 & Spacelike: $S_{X}=\frac{c_{\text{AdS}}}{3}\log\left(\frac{L}{\pi\epsilon}\sin\left(\frac{\pi X}{L}\right)\right)$ & $\delta S_{X}=0$\tabularnewline
\hline 
\multirow{2}{*}{Finite temperature} & Timelike: $S_{T}=\frac{c_{\text{AdS}}}{3}\log\left(\frac{\beta}{\pi\epsilon}\sinh\frac{\pi T}{\epsilon}\right)+\frac{ic_{\text{AdS}}\pi}{6}$ & $\delta S_{T}=0$\tabularnewline
\cline{2-3} \cline{3-3} 
 & Spacelike: $S_{X}=\frac{c_{\text{AdS}}}{3}\log\left(\frac{\beta}{\pi\epsilon}\sinh\frac{\pi X}{\epsilon}\right)$ & $\delta S_{X}=-\mu\frac{\pi^{4}c_{\text{AdS}}^{2}}{9\beta^{3}}X\coth\left(\frac{\pi X}{\beta}\right)$\tabularnewline
\hline 
\end{tabular}
\par\end{centering}
\caption{\label{tab:tab2}The leading correction to the spacelike/timelike EE
caused by the $T\bar{T}$ deformation in finite size/temperature system.}
\end{table}

In the above field-theoretic results, the leading $T\bar{T}$ correction
to the timelike EE exists in finite size systems
but vanishes in finite temperature systems, while the leading $T\bar{T}$
correction to the spacelike EE exhibits the opposite
behavior, as shown in Table \ref{tab:tab2}, 
which is in perfect agreement with our prediction in \cite{Wang:2018jva}. 
Meanwhile, the leading
$T\bar{T}$ correction to the general entanglement entropy always
exists in both finite size systems and finite temperature systems.
Therefore, the general entanglement entropy is the right measure to mark
the deformations.

\section{Gravity duals}

\label{sec:Gravity-dual}
It is illuminating to study the spacelike/timelike EE and its corresponding
gravity dual in the context of cAdS/dCFT correspondence. In this section,
we will demonstrate that the distance of the geodesic in the cutoff-AdS
precisely matches the general entanglement entropy in the $T\bar{T}$
deformed $\text{CFT}_{2}$. We will only compute the geodesic length between two boundary points, which does not prove the existence of  the geometric bulk dual of the general EE. However, the bulk dual of the timelike EE has been explicitly studied in \cite{Li:2022tsv}.
\begin{figure}[H]
\begin{centering}
\includegraphics[width=0.25\textwidth]{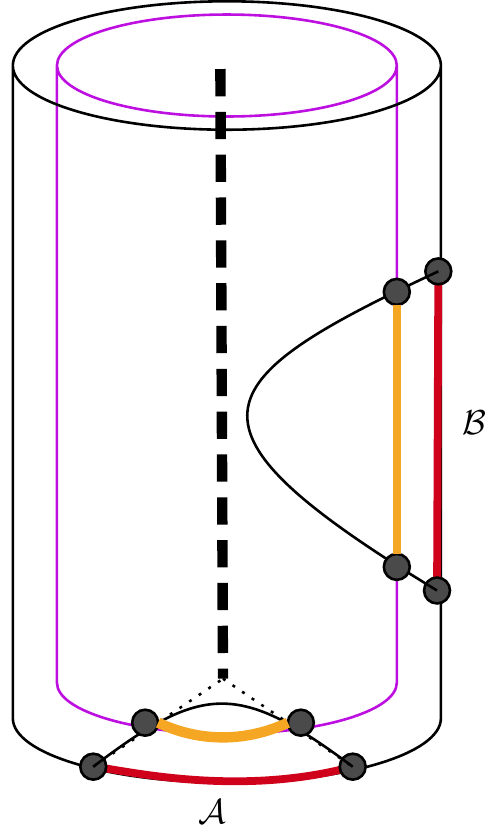} 
\par\end{centering}
\caption{The figure illustrates a Euclidean cylindrical manifold with a black cylinder representing the original boundary and a violet cylinder representing the cut-off boundary. Two red lines denote an interval $\mathcal{A}$ in the compact direction and an interval $\mathcal{B}$ in the non-compact direction. Two orange lines indicate their corresponding intervals in the cut-off boundary.}
\label{fig:figure2}
\end{figure}
Before that, we first provide a physical interpretation of our previous conjecture by 
introducing a Euclidean cylindrical manifold, as depicted in Figure \ref{fig:figure2}, 
which features one compact and one non-compact direction. 
The compact direction is always described by a native parameter, 
either the inverse temperature $\beta$ in the finite temperature system or 
the total length $L$ in the finite size system. As a result, the interval $\mathcal{A}$ in 
the compact direction is independent of the cut-off boundary 
and the geodesic anchored on $\partial \mathcal{A}$ remains unchanged. 
The interval $\mathcal{A}$ in the compact direction is independent of the choice of the cut-off boundary, and the geodesic anchored on $\partial \mathcal{A}$ will not change. 
In contrast, the interval $\mathcal{B}$ in the non-compact direction will vary with the cut-off boundary, which leads to a change in the geodesic anchored on $\partial \mathcal{B}$. Therefore, in the finite temperature system, the temporal direction is compact and its timelike EE remains unaffected by the $T\bar{T}$ deformation; in the finite size system, the spatial direction is compact and its spacelike EE remains unchanged under the $T\bar{T}$ deformation.
 
\subsection{BTZ black hole}

\label{subsec:BTZ-black-hole}

Consider a BTZ black hole that is described by

\begin{equation}
ds^{2}=-\frac{r^{2}-r_{+}^{2}}{R_{\text{AdS}}^{2}}dt^{2}+\frac{R_{\text{AdS}}^{2}}{r^{2}-r_{+}^{2}}dr^{2}+\frac{r^{2}}{R_{\text{AdS}}^{2}}dx^{2},\label{eq:BTZmetric}
\end{equation}
where $r_{+}$ is the radius of event horizon, and $t$ is the compact
temporal direction $t\sim t+i\beta$. It is well-known that the finite
temperature $\text{CFT}_{2}$ is dual to a BTZ black hole with the
same temperature 
\begin{equation}
\beta^{-1}=\frac{r_{+}}{2\pi R_{\text{AdS}}^{2}},\label{eq:temp-1}
\end{equation}
In the cAdS/dCFT correspondence, the $T\bar{T}$ deformed CFT at finite
temperature is dual to a BTZ black hole with the radial cutoff $r_{c}$,
\begin{equation}
r_{c}^{2}=\frac{6R_{\text{AdS}}^{4}}{\pi c_{\text{AdS}}\mu}.\label{eq:cutoff-1}
\end{equation}
Taking $r=r_{c}$ and multiplying a factor $R_{\text{AdS}}^{2}/(r_{c}^{2}-r_{+}^{2})$
in the BTZ black hole metric, the metric of the cutoff boundary, where
the $T\bar{T}$ deformed CFT lives, reads
\begin{equation}
ds^{2}=-dt^{2}+\frac{dx^{2}}{1-r_{+}^{2}/r_{c}^{2}}.
\end{equation}
Notice that the timelike interval on the boundary keeps invariant,
because the compact temporal direction should be physical. This is
the reason that the timelike EE does not receive
a correction from the $T\bar{T}$ deformation. To see this, for
an interval $A$, with the timelike width $t_{0}$ and the spacelike
width $x_{0}$, in  the $T\bar{T}$ deformed CFT, we calculate the
general entanglement entropy with the RT formula. Performing
the following coordinates transformation,
\begin{eqnarray}
u & = & \sqrt{1-\frac{r_{+}^{2}}{r^{2}}}\cosh\left(\frac{r_{+}t}{R_{\text{AdS}}^{2}}\right)\exp\left(\frac{r_{+}x}{R_{\text{AdS}}^{2}}\right),\nonumber \\
v & = & \sqrt{1-\frac{r_{+}^{2}}{r^{2}}}\sinh\left(\frac{r_{+}t}{R_{\text{AdS}}^{2}}\right)\exp\left(\frac{r_{+}x}{R_{\text{AdS}}^{2}}\right),\\
z & = & \frac{r_{+}}{r}\exp\left(\frac{r_{+}x}{R_{\text{AdS}}^{2}}\right),\nonumber 
\end{eqnarray}
the metric (\ref{eq:BTZmetric}) becomes
\begin{equation}
ds^{2}=\frac{R_{\text{AdS}}^{2}}{z^{2}}\left(du^{2}-dv^{2}+dz^{2}\right).
\end{equation}
Two boundary points $\partial A=\left\{ \left(u_{1},v_{1},z_{1}\right),\left(u_{2},v_{2},z_{2}\right)\right\} $
could be written as 
\begin{equation}
\begin{cases}
u_{1}=\sqrt{1-\frac{r_{+}^{2}}{r_{c}^{2}}}, & u_{2}=\sqrt{1-\frac{r_{+}^{2}}{r_{c}^{2}}}\cosh\left(\frac{r_{+}t_{0}}{R_{\text{AdS}}^{2}}\right)\exp\left(\frac{r_{+}x_{0}}{R_{\text{AdS}}^{2}}\sqrt{1-\frac{r_{+}^{2}}{r_{c}^{2}}}\right),\\
v_{1}=0, & v_{2}=\sqrt{1-\frac{r_{+}^{2}}{r_{c}^{2}}}\sinh\left(\frac{r_{+}t_{0}}{R_{\text{AdS}}^{2}}\right)\exp\left(\frac{r_{+}x_{0}}{R_{\text{AdS}}^{2}}\sqrt{1-\frac{r_{+}^{2}}{r_{c}^{2}}}\right),\\
z_{1}=\frac{r_{+}}{r_{c}}, & z_{2}=\frac{r_{+}}{r_{c}}\exp\left(\frac{r_{+}x_{0}}{R_{\text{AdS}}^{2}}\sqrt{1-r_{+}^{2}/r_{c}^{2}}\right).
\end{cases}
\end{equation}
Therefore, one can determine the length of the geodesic $\gamma_{A}$
to be 
\begin{eqnarray}
L_{\gamma_{A}} & = & R_{\text{AdS}}\,\mathrm{arcosh}\left[1+\frac{\left(u_{2}-u_{1}\right)^{2}-\left(v_{2}-v_{1}\right)^{2}+\left(z_{2}-z_{1}\right)^{2}}{2z_{1}z_{2}}\right]\nonumber \\
 & = & R_{\text{AdS}}\,\mathrm{arcosh}\left[\frac{\left(r_{+}^{2}/r_{c}^{2}-1\right)\cosh\left(\frac{r_{+}t_{0}}{R_{\text{AdS}}^{2}}\right)+\cosh\left(\frac{r_{+}x_{0}}{R_{\text{AdS}}^{2}}\sqrt{1-r_{+}^{2}/r_{c}^{2}}\right)}{r_{+}^{2}/r_{c}^{2}}\right]\nonumber \\
 & = & 2R_{\text{AdS}}\log\left[\frac{r_{c}}{r_{+}}\sqrt{2\cosh\left(\frac{r_{+}x_{0}}{R_{\text{AdS}}^{2}}\right)-2\cosh\left(\frac{r_{+}t_{0}}{R_{\text{AdS}}^{2}}\right)}\right]-\frac{r_{+}^{2}}{r_{c}^{2}}\frac{\frac{r_{+}x_{0}}{2R_{\text{AdS}}}\sinh(\frac{r_{+}x_{0}}{R_{\text{AdS}}^{2}})}{\cosh(\frac{r_{+}x_{0}}{R_{\text{AdS}}^{2}})-\cosh(\frac{r_{+}t_{0}}{R_{\text{AdS}}^{2}})}\nonumber \\
 &  & +\frac{r_{+}^{2}}{r_{c}^{2}}\frac{R_{\text{AdS}}\,\cosh(\frac{r_{+}t_{0}}{R_{\text{AdS}}^{2}})}{\cosh(\frac{r_{+}x_{0}}{R_{\text{AdS}}^{2}})-\cosh(\frac{r_{+}t_{0}}{R_{\text{AdS}}^{2}})}+O\left(\frac{r_{+}^{3}}{r_{c}^{3}}\right).
\end{eqnarray}
Notice that the first term in the last line should not be thought
of a leading correction caused by $T\bar{T}$ flow. To see this point,
replacing the metric of the cutoff boundary with $ds^{2}=-dt^{2}+dx^{2}$,
one can compute the length of the geodesic $\gamma_{A}$:
\begin{eqnarray}
L_{\gamma_{A}} & = & R_{\text{AdS}}\,\mathrm{arcosh}\left[\frac{\left(r_{+}^{2}/r_{c}^{2}-1\right)\cosh\left(\frac{r_{+}t_{0}}{R_{\text{AdS}}^{2}}\right)+\cosh\left(\frac{r_{+}x_{0}}{R_{\text{AdS}}^{2}}\right)}{r_{+}^{2}/r_{c}^{2}}\right]\nonumber \\
 & = & 2R_{\text{AdS}}\log\left[\frac{r_{c}}{r_{+}}\sqrt{2\cosh\left(\frac{r_{+}x_{0}}{R_{\text{AdS}}^{2}}\right)-2\cosh\left(\frac{r_{+}t_{0}}{R_{\text{AdS}}^{2}}\right)}\right]\nonumber \\
 &  & +\frac{r_{+}^{2}}{r_{c}^{2}}\frac{R_{\text{AdS}}\cosh(\frac{r_{+}t_{0}}{R_{\text{AdS}}^{2}})}{\cosh(\frac{r_{+}x_{0}}{R_{\text{AdS}}^{2}})-\cosh(\frac{r_{+}t_{0}}{R_{\text{AdS}}^{2}})}+O\left(\frac{r_{+}^{3}}{r_{c}^{3}}\right).
\end{eqnarray}
Therefore, utilizing equations (\ref{eq:temp-1}) and (\ref{eq:cutoff-1})
and identifying $\epsilon^{2}=\pi c_{\text{AdS}}\mu/6$, the right estimation of
the general entanglement entropy corrected by $T\bar{T}$ deformation
is 
\begin{equation}
\frac{L_{\gamma_{A}}}{4G_{N}}\sim\frac{R_{\text{AdS}}}{2G_{N}}\log\left[\frac{\beta}{2\pi\epsilon}\sqrt{2\cosh\left(\frac{2\pi x_{0}}{\beta}\right)-2\cosh\left(\frac{2\pi t_{0}}{\beta}\right)}\right]-\frac{R_{\text{AdS}}}{2G_{N}}\frac{\pi^{3}c_{\text{AdS}}\mu}{3\beta^{2}}\frac{\frac{\pi x_{0}}{\beta}\sinh(\frac{2\pi x_{0}}{\beta})}{\cosh(\frac{2\pi x_{0}}{\beta})-\cosh(\frac{2\pi t_{0}}{\beta})},
\end{equation}
that is 
\begin{equation}
S_{A}^{(\mu)}=\frac{c_{\text{AdS}}}{3}\log\left[\frac{\beta}{2\pi\epsilon}\sqrt{2\cosh\left(\frac{2\pi x_{0}}{\beta}\right)-2\cosh\left(\frac{2\pi t_{0}}{\beta}\right)}\right]-\mu\frac{\pi^{4}c_{\text{AdS}}^{2}}{9\beta^{3}}\frac{x_{0}\sinh(\frac{2\pi x_{0}}{\beta})}{\cosh(\frac{2\pi x_{0}}{\beta})-\cosh(\frac{2\pi t_{0}}{\beta})}
\end{equation}
This precisely matches the field theoretic result (\ref{eq:temp-gen-ee}).

\subsection{Global $\text{AdS}_{3}$}

\label{subsec:Global}

Consider the finite size $\text{CFT}_{2}$ at zero temperature that
is dual to the global $\text{AdS}_{3}$. The metric of the global
$\text{AdS}_{3}$ is given by
\begin{equation}
ds^{2}=R_{\text{AdS}}^{2}\left(-\cosh^{2}\rho\,d\theta^{2}+d\rho^{2}+\sinh^{2}\rho\,d\phi^{2}\right),\label{eq:adsmetric}
\end{equation}
where $\phi$ is the compact spatial direction $\phi\sim\phi+2\pi$.
In the cAdS/dCFT correspondence, the $T\bar{T}$ deformed CFT lives
at the radial cutoff $\rho_{c}$,
\begin{equation}
\cosh^{2}\rho_{c}=\frac{3L^{2}}{2\mu\pi^{3}c_{\text{AdS}}},\label{eq:global-cutoff}
\end{equation}
where the total length of the boundary circle is $L$, and the metric
of the cutoff boundary reads
\begin{equation}
ds^{2}=-\coth^{2}\rho_{c}\,d\theta^{2}+d\phi^{2}.
\end{equation}
Intriguingly, the timelike interval in the boundary is indeed changed
by $T\bar{T}$ deformation, which means that timelike EE
in the finite size system will be corrected by $T\bar{T}$
deformation. By choosing an interval $A$ with the timelike width
$t_{0}=\frac{L\theta_{0}}{2\pi}$ and the spacelike width $x_{0}=\frac{L\phi_{0}}{2\pi}$,
and performing the following coordinates transformation,
\begin{eqnarray}
u & = & \tanh\rho\cos\left(\theta\right)\exp\left(i\phi\right),\nonumber \\
v & = & \tanh\rho\sin\left(\theta\right)\exp\left(i\phi\right),\\
z & = & \frac{1}{\cosh\rho}\exp\left(i\phi\right),\nonumber 
\end{eqnarray}
the metric (\ref{eq:adsmetric}) becomes
\begin{equation}
ds^{2}=\frac{R_{\text{AdS}}^{2}}{z^{2}}\left(du^{2}+dv^{2}+dz^{2}\right).
\end{equation}
The two boundary points $\partial A=\left\{ \left(u_{1},v_{1},z_{1}\right),\left(u_{2},v_{2},z_{2}\right)\right\} $
could be written as 
\begin{equation}
\begin{cases}
u_{1}=\tanh\rho_{c}, & u_{2}=\tanh\rho_{c}\cos\left(\frac{2\pi x_{0}}{L}\right)\exp\left(i\frac{2\pi t_{0}}{L}\tanh\rho_{c}\right),\\
v_{1}=0, & v_{2}=\tanh\rho_{c}\sin\left(\frac{2\pi x_{0}}{L}\right)\exp\left(i\frac{2\pi t_{0}}{L}\tanh\rho_{c}\right),\\
z_{1}=\frac{1}{\cosh\rho_{c}}, & z_{2}=\frac{1}{\cosh\rho_{c}}\exp\left(i\frac{2\pi t_{0}}{L}\tanh\rho_{c}\right),
\end{cases}
\end{equation}
and the distance of the geodesic $\gamma_{A}$, anchored on $\partial A$,
could be captured by 

\begin{eqnarray}
L_{\gamma_{A}} & = & R_{\text{AdS}}\,\mathrm{arcosh}\left[\frac{\cos\left(\frac{2\pi t_{0}}{L}\tanh\rho_{c}\right)+\left(\cosh^{-2}\rho_{c}-1\right)\cos\left(\frac{2\pi x_{0}}{L}\right)}{\cosh^{-2}\rho_{c}}\right]\nonumber \\
 & = & 2R_{\text{AdS}}\log\left[\cosh\rho_{c}\sqrt{2\cos\left(\frac{2\pi t_{0}}{L}\right)-2\cos\left(\frac{2\pi x_{0}}{L}\right)}\right]+\frac{\frac{R_{\text{AdS}}}{\cosh^{2}\rho_{c}}\frac{\pi t_{0}}{L}\sin\left(\frac{2\pi t_{0}}{L}\right)}{\cos\left(\frac{2\pi t_{0}}{L}\right)-\cos\left(\frac{2\pi x_{0}}{L}\right)}\nonumber \\
 &  & +\frac{R_{\text{AdS}}}{\cosh^{2}\rho_{c}}\frac{\cos\left(\frac{2\pi t_{0}}{L}\right)}{\cos\left(\frac{2\pi t_{0}}{L}\right)-\cos\left(\frac{2\pi x_{0}}{L}\right)}+O\left(\frac{1}{\cosh^{3}\rho_{c}}\right)\\
 & \sim & 2R_{\text{AdS}}\log\left[\cosh\rho_{c}\sqrt{2\cos\left(\frac{2\pi t_{0}}{L}\right)-2\cos\left(\frac{2\pi x_{0}}{L}\right)}\right]+\frac{\frac{R_{\text{AdS}}}{\cosh^{2}\rho_{c}}\frac{\pi t_{0}}{L}\sin\left(\frac{2\pi t_{0}}{L}\right)}{\cos\left(\frac{2\pi t_{0}}{L}\right)-\cos\left(\frac{2\pi x_{0}}{L}\right)}.\label{eq:global-geo}
\end{eqnarray}
Substituting equation (\ref{eq:global-cutoff}) into equation (\ref{eq:global-geo}),
and identifying $\epsilon^{2}=\pi c\mu/6$, one obtains
\begin{eqnarray}
L_{\gamma_{A}} & \sim & 2R_{\text{AdS}}\log\left[\frac{L}{2\pi\epsilon}\sqrt{2\cos\left(\frac{2\pi t_{0}}{L}\right)-2\cos\left(\frac{2\pi x_{0}}{L}\right)}\right]+\frac{2\mu\pi^{4}c_{\text{AdS}}}{3L^{3}}\frac{R_{\text{AdS}}t_{0}\sin\left(\frac{2\pi t_{0}}{L}\right)}{\cos\left(\frac{2\pi t_{0}}{L}\right)-\cos\left(\frac{2\pi x_{0}}{L}\right)}.
\end{eqnarray}
With the RT formula, one finds the following estimation
of the general entanglement entropy:
\begin{eqnarray}
S_{A}^{(\mu)} & = & \frac{c_{\text{AdS}}}{3}\log\left[\frac{L}{2\pi\epsilon}\sqrt{2\cos\left(\frac{2\pi t_{0}}{L}\right)-2\cos\left(\frac{2\pi x_{0}}{L}\right)}\right]+\frac{\mu\pi^{4}c_{\text{AdS}}^{2}}{9L^{3}}\frac{t_{0}\sin\left(\frac{2\pi t_{0}}{L}\right)}{\cos\left(\frac{2\pi t_{0}}{L}\right)-\cos\left(\frac{2\pi x_{0}}{L}\right)},
\end{eqnarray}
which is a perfect match with the field theoretic result (\ref{eq:field-size-timelike}).

%\subsection{General entanglement entropy in dS/CFT}
%
%\label{subsec:Comments-on-dS/CFT}
%
%Let us consider the static coordinates of $\text{dS}_{3}$ spacetime
%and the metric takes the form
%\begin{equation}
%ds^{2}=-\left(R_{\text{dS}}^{2}-r^{2}\right)d\theta^{2}+\left(1-\frac{r^{2}}{R_{\text{dS}}^{2}}\right)^{-1}dr^{2}+r^{2}d\phi^{2},
%\end{equation}
%Performing the coordinates transformations $r=R_{\text{dS}}\cos\varrho$,
%the $\text{dS}_{3}$ spacetime metric reads
%\begin{equation}
%ds^{2}=R_{\text{dS}}^{2}\left(\sin^{2}\varrho\,d\theta^{2}-d\varrho^{2}+\cos^{2}\varrho\,d\phi^{2}\right),
%\end{equation}
%This matches the global $\text{AdS}_{3}$ spacetime (\ref{eq:adsmetric})
%via the analytical continuation $R_{\text{dS}}=iR_{\text{AdS}}$,
%$\varrho=i\rho$ and exchanging the temporal direction with the spatial
%direction $\theta\leftrightarrow\phi$. Therefore, the general entanglement
%entropy in $\text{dS}_{3}/\text{CFT}_{2}$ is 
%\begin{equation}
%S_{A}=-i\frac{c_{\text{dS}}}{3}\log\left[\frac{L}{2\pi\epsilon}\sqrt{2\cos\left(\frac{2\pi x_{0}}{L}\right)-2\cos\left(\frac{2\pi t_{0}}{L}\right)}\right].
%\end{equation}
%As a consistent check, setting $t_{0}=0$ and $L=2\pi$, one obtains
%\begin{equation}
%S_{A}=-i\frac{c_{\text{dS}}}{3}\log\left[\frac{2}{\epsilon}\sin\left(\frac{x_{0}}{2}\right)\right]+\frac{\pi c_{\text{dS}}}{6},
%\end{equation}
%which is exactly the result in \cite{Doi:2022iyj}.

\section{Conclusions}

\label{Sec:Conc}

The $T\bar{T}$ deformation has been widely studied in recent years,
due to its integrability in 2$d$ CFTs and its applications in holography.
In a previous work, we had amazingly found that the usual spacelike
entanglement entropy alone is not sufficient to fully mark
the $T\bar{T}$ deformation and that a timelike
entanglement entropy must be introduced. The conventional spacelike entanglement entropy fails to fully capture the entangling properties, as it remains unchanged between the undeformed finite-size CFT and the $T\bar{T}$ deformed one. In this paper, we show that, complementarily, the timelike entanglement entropy could distinguish the undeformed finite-size CFT and the $T\bar{T}$ deformed one. In the context of cAdS/dCFT correspondence, we affirm the indispensability of the timelike entanglement entropy, as it proves crucial in differentiating between the undeformed CFT and the $T\bar{T}$ deformed CFT. Our main results have been shown in Table \ref{tab:tab2}.

In this paper, our findings heavily rely on the AdS/CFT correspondence. It is equally crucial to investigate the timelike entanglement entropy within the framework of the dS/CFT correspondence. Moreover, the physical interpretation of the timelike entanglement entropy remains obscure, making explicit studies on this topic of utmost importance. Consequently, the timelike entanglement entropy may hold significant potential in enhancing our comprehension of black hole information and the emergence of spacetime geometry from entanglement.

%In this paper, we proposed a general entanglement entropy that consists of both spacelike and timelike components. We demonstrated that indeed the leading $T\bar{T}$ correction to the general entanglement entropy exists in both finite size systems and finite temperature systems. We also utilized the holographic method to evaluate the general entanglement entropy. 
\vspace{3ex}

\noindent {\bf Acknowledgements} 
This work is supported in part by NSFC (Grant No. 12275183, 12275184, 12105191 and 11875196). HW is supported by the International Visiting Program for Excellent Young Scholars of Sichuan University.

\appendix
%dummy comment inserted by tex2lyx to ensure that this paragraph is not empty%dummy comment inserted by tex2lyx to ensure that this paragraph is not empty

\bibliographystyle{unsrturl}
\bibliography{gEE_v3}

\begin{thebibliography}{10}

\bibitem{Wang:2018jva}
Peng Wang, Houwen Wu, and Haitang Yang.
\newblock {Fix the dual geometries of $T\bar{T}$ deformed CFT$_2$ and highly
  excited states of CFT$_2$}.
\newblock {\em Eur. Phys. J. C}, 80(12):1117, 2020.
\newblock \href {http://arxiv.org/abs/1811.07758} {\path{arXiv:1811.07758}},
  \href {https://doi.org/10.1140/epjc/s10052-020-08680-7}
  {\path{doi:10.1140/epjc/s10052-020-08680-7}}.

\bibitem{Zamolodchikov:2004ce}
Alexander~B. Zamolodchikov.
\newblock {Expectation value of composite field T anti-T in two-dimensional
  quantum field theory}.
\newblock 1 2004.
\newblock \href {http://arxiv.org/abs/hep-th/0401146}
  {\path{arXiv:hep-th/0401146}}.

\bibitem{Cavaglia:2016oda}
Andrea Cavagli\`a, Stefano Negro, Istv\'an~M. Sz\'ecs\'enyi, and Roberto Tateo.
\newblock {$T \bar{T}$-deformed 2D Quantum Field Theories}.
\newblock {\em JHEP}, 10:112, 2016.
\newblock \href {http://arxiv.org/abs/1608.05534} {\path{arXiv:1608.05534}},
  \href {https://doi.org/10.1007/JHEP10(2016)112}
  {\path{doi:10.1007/JHEP10(2016)112}}.

\bibitem{Smirnov:2016lqw}
F.~A. Smirnov and A.~B. Zamolodchikov.
\newblock {On space of integrable quantum field theories}.
\newblock {\em Nucl. Phys. B}, 915:363--383, 2017.
\newblock \href {http://arxiv.org/abs/1608.05499} {\path{arXiv:1608.05499}},
  \href {https://doi.org/10.1016/j.nuclphysb.2016.12.014}
  {\path{doi:10.1016/j.nuclphysb.2016.12.014}}.

\bibitem{McGough:2016lol}
Lauren McGough, M\'ark Mezei, and Herman Verlinde.
\newblock {Moving the CFT into the bulk with $ T\overline{T} $}.
\newblock {\em JHEP}, 04:010, 2018.
\newblock \href {http://arxiv.org/abs/1611.03470} {\path{arXiv:1611.03470}},
  \href {https://doi.org/10.1007/JHEP04(2018)010}
  {\path{doi:10.1007/JHEP04(2018)010}}.

\bibitem{Maldacena:1997re}
Juan~Martin Maldacena.
\newblock {The Large N limit of superconformal field theories and
  supergravity}.
\newblock {\em Adv. Theor. Math. Phys.}, 2:231--252, 1998.
\newblock \href {http://arxiv.org/abs/hep-th/9711200}
  {\path{arXiv:hep-th/9711200}}, \href
  {https://doi.org/10.1023/A:1026654312961}
  {\path{doi:10.1023/A:1026654312961}}.

\bibitem{He:2019vzf}
Song He and Hongfei Shu.
\newblock {Correlation functions, entanglement and chaos in the $
  T\overline{T}/J\overline{T} $-deformed CFTs}.
\newblock {\em JHEP}, 02:088, 2020.
\newblock \href {http://arxiv.org/abs/1907.12603} {\path{arXiv:1907.12603}},
  \href {https://doi.org/10.1007/JHEP02(2020)088}
  {\path{doi:10.1007/JHEP02(2020)088}}.

\bibitem{He:2020qcs}
Song He.
\newblock {Note on higher-point correlation functions of the $T\bar T$ or
  $J\bar T$ deformed CFTs}.
\newblock {\em Sci. China Phys. Mech. Astron.}, 64(9):291011, 2021.
\newblock \href {http://arxiv.org/abs/2012.06202} {\path{arXiv:2012.06202}},
  \href {https://doi.org/10.1007/s11433-021-1741-1}
  {\path{doi:10.1007/s11433-021-1741-1}}.

\bibitem{He:2022xkh}
Song He, Zhang-Cheng Liu, and Yuan Sun.
\newblock {Entanglement entropy and modular Hamiltonian of free fermion with
  deformations on a torus}.
\newblock {\em JHEP}, 09:247, 2022.
\newblock \href {http://arxiv.org/abs/2207.06308} {\path{arXiv:2207.06308}},
  \href {https://doi.org/10.1007/JHEP09(2022)247}
  {\path{doi:10.1007/JHEP09(2022)247}}.

\bibitem{Calabrese:2004eu}
Pasquale Calabrese and John~L. Cardy.
\newblock {Entanglement entropy and quantum field theory}.
\newblock {\em J. Stat. Mech.}, 0406:P06002, 2004.
\newblock \href {http://arxiv.org/abs/hep-th/0405152}
  {\path{arXiv:hep-th/0405152}}, \href
  {https://doi.org/10.1088/1742-5468/2004/06/P06002}
  {\path{doi:10.1088/1742-5468/2004/06/P06002}}.

\bibitem{Calabrese:2009qy}
Pasquale Calabrese and John Cardy.
\newblock {Entanglement entropy and conformal field theory}.
\newblock {\em J. Phys. A}, 42:504005, 2009.
\newblock \href {http://arxiv.org/abs/0905.4013} {\path{arXiv:0905.4013}},
  \href {https://doi.org/10.1088/1751-8113/42/50/504005}
  {\path{doi:10.1088/1751-8113/42/50/504005}}.

\bibitem{Chen:2018eqk}
Bin Chen, Lin Chen, and Peng-Xiang Hao.
\newblock {Entanglement entropy in $T\overline{T}$-deformed CFT}.
\newblock {\em Phys. Rev. D}, 98(8):086025, 2018.
\newblock \href {http://arxiv.org/abs/1807.08293} {\path{arXiv:1807.08293}},
  \href {https://doi.org/10.1103/PhysRevD.98.086025}
  {\path{doi:10.1103/PhysRevD.98.086025}}.

\bibitem{Doi:2022iyj}
Kazuki Doi, Jonathan Harper, Ali Mollabashi, Tadashi Takayanagi, and Yusuke
  Taki.
\newblock {Pseudoentropy in dS/CFT and Timelike Entanglement Entropy}.
\newblock {\em Phys. Rev. Lett.}, 130(3):031601, 2023.
\newblock \href {http://arxiv.org/abs/2210.09457} {\path{arXiv:2210.09457}},
  \href {https://doi.org/10.1103/PhysRevLett.130.031601}
  {\path{doi:10.1103/PhysRevLett.130.031601}}.

\bibitem{Narayan:2022afv}
K.~Narayan.
\newblock {de Sitter space, extremal surfaces and ''time-entanglement''}.
\newblock 10 2022.
\newblock \href {http://arxiv.org/abs/2210.12963} {\path{arXiv:2210.12963}}.

\bibitem{Doi:2023zaf}
Kazuki Doi, Jonathan Harper, Ali Mollabashi, Tadashi Takayanagi, and Yusuke
  Taki.
\newblock {Timelike entanglement entropy}.
\newblock 2 2023.
\newblock \href {http://arxiv.org/abs/2302.11695} {\path{arXiv:2302.11695}}.

\bibitem{Narayan:2023ebn}
K.~Narayan and Hitesh~K. Saini.
\newblock {Notes on time entanglement and pseudo-entropy}.
\newblock 3 2023.
\newblock \href {http://arxiv.org/abs/2303.01307} {\path{arXiv:2303.01307}}.

\bibitem{Ryu:2006bv}
Shinsei Ryu and Tadashi Takayanagi.
\newblock {Holographic derivation of entanglement entropy from AdS/CFT}.
\newblock {\em Phys. Rev. Lett.}, 96:181602, 2006.
\newblock \href {http://arxiv.org/abs/hep-th/0603001}
  {\path{arXiv:hep-th/0603001}}, \href
  {https://doi.org/10.1103/PhysRevLett.96.181602}
  {\path{doi:10.1103/PhysRevLett.96.181602}}.

\bibitem{Narayan:2015vda}
K.~Narayan.
\newblock {Extremal surfaces in de Sitter spacetime}.
\newblock {\em Phys. Rev. D}, 91(12):126011, 2015.
\newblock \href {http://arxiv.org/abs/1501.03019} {\path{arXiv:1501.03019}},
  \href {https://doi.org/10.1103/PhysRevD.91.126011}
  {\path{doi:10.1103/PhysRevD.91.126011}}.

\bibitem{Narayan:2015oka}
K.~Narayan.
\newblock {de Sitter space and extremal surfaces for spheres}.
\newblock {\em Phys. Lett. B}, 753:308--314, 2016.
\newblock \href {http://arxiv.org/abs/1504.07430} {\path{arXiv:1504.07430}},
  \href {https://doi.org/10.1016/j.physletb.2015.12.019}
  {\path{doi:10.1016/j.physletb.2015.12.019}}.

\bibitem{Narayan:2016xwq}
K.~Narayan.
\newblock {On $dS_4$ extremal surfaces and entanglement entropy in some ghost
  CFTs}.
\newblock {\em Phys. Rev. D}, 94(4):046001, 2016.
\newblock \href {http://arxiv.org/abs/1602.06505} {\path{arXiv:1602.06505}},
  \href {https://doi.org/10.1103/PhysRevD.94.046001}
  {\path{doi:10.1103/PhysRevD.94.046001}}.

\bibitem{Li:2022tsv}
Ze~Li, Zi-Qing Xiao, and Run-Qiu Yang.
\newblock {On holographic time-like entanglement entropy}.
\newblock 11 2022.
\newblock \href {http://arxiv.org/abs/2211.14883} {\path{arXiv:2211.14883}}.

\end{thebibliography}

\end{document}